\documentclass[10pt,draftcls,onecolumn]{IEEEtran}
\usepackage{graphicx}
\usepackage{amsmath}
\usepackage{amssymb}
\usepackage{epstopdf}
\usepackage{mathrsfs}
\usepackage{subfigure}
\usepackage{caption}
\usepackage{cite}
\usepackage{algorithm}
\usepackage{algorithmic}
\newcommand{\nonumSTATE}{\item[]}

\begin{document}
\title{Low Rank and Sparsity Analysis Applied to Speech Enhancement via Online Estimated Dictionary}
\author{Pengfei~Sun,~and~Jun~Qin} 


\maketitle
\begin{abstract}
In this letter, we propose an online estimated dictionary based single-channel speech enhancement algorithm, which focuses on low-rank and sparse matrix decomposition. In the proposed algorithm, a noisy speech spectrogram can be decomposed into low rank background noise components and an activation of the online speech dictionary, on which both low-rank and sparsity constraints are imposed. This decomposition takes the advantage of local estimated exemplar's high expressiveness on speech components and also accommodates nonstationary background noise. The local dictionary can be obtained through estimating the speech presence probability (SPP) by applying Expectation-Maximal algorithm, in which a generalized Gamma prior for speech magnitude spectrum is used. The proposed algorithm is evaluated using signal-to-distortion ratio (SDR), and perceptual evaluation of speech quality (PESQ). The results show that the proposed algorithm achieves significant improvements at various SNRs when compared to four other speech enhancement algorithms, including improved Karhunen-Loeve transform (KLT) approach, SPP based MMSE (MMSE-SPP), NMF based RPCA (NMF-RPCA), and RPCA.
\end{abstract}
\IEEEpeerreviewmaketitle
\section{Introduction}
\IEEEPARstart{S}{ingle} channel speech enhancement is a key issue in speech processing, aiming to improve the performance of speech communication systems in noisy environments. Recently-developed robust principal component analysis (RPCA) has been shown effective in separating the speech components from background noise \cite{sun2014novel,elhamifar2013sparse, wu2014theory}. This approach decomposes the spectrogram matrix as the sum of a sparse matrix and a low-rank matrix, representing speech and noise, respectively\cite{sun2015speech}. Based on the observation that unpredictable background noise is often less spectrally diverse than the foreground speech, it regards that sparse matrix represents the speech components while noise is included in the low rank matrix \cite{bouzid2016speech}. Because the activation of the low-rank components can be temporally variable, this unsupervised decomposition can accommodate nonstationary noise \cite{sprechmann2012real}. However, the targeted speech may also contain low rank components described by limited spectral bases, and these low rank speech components can be wrongly decomposed into noise matrix. Thus unsupervised decomposition may not provide accurate separation of foreground and background in certain circumstance, e.g., nonstationary noise\cite{mohammadiha2013supervised}.

To avoid simply using sparsity to characterize speech spectrogram, incorporating the knowledge about the likely form of the targeted speech has been introduced \cite{chen2013speech, mohammadiha2013supervised}. Using nonnegative matrix factorization (NMF) technique, pre-learned explicit speech dictionary \cite{duan2012speech, sprechmann2014supervised} based on large dataset can improve the performance of RPCA. The advantage of the offline-trained global dictionary based RPCA approach is its ability to maintain the flexibility to distinguish speech and noise, and also avoid wrongly decomposing the low-rank speech components into noise subspace. However, offline-learned global basis spectra may either lead to a non-'sparse' activation matrix or wrongly interpret background noise as speech. This is intrinsically due to overfitting caused by limited local speech eigenvectors and shared basis spectra between speech and noise \cite{duan2012speech}. Recent study \cite{mohammadiha2014single} indicates that exemplar based dictionary can be more effective on covering speech spectra convex hull. In addition, online trained dictionary \cite{duan2012speech} presents a solution to achieve higher noise reduction whereas lower speech distortion. To exploit the advantages of both, a natural idea is to develop an online learned exemplar-structure dictionary.   

In this study, we propose an online estimated dictionary based low rank and sparse decomposition (LDLSD) algorithm. The developed local dictionary estimation inherits the merit of both exemplar's efficient explanation on the speech manifold and adaptive speech subspace that alleviating the immediate speech distortion \cite{elhamifar2013sparse}. In LDLSD algorithm, a semi-supervised Expectation-Maximum (EM) algorithm based on a generalized Gamma speech distribution model is used to calculate the speech presence probability (SPP), and further obtain the local estimated exemplar. Moreover, the activation matrix of speech is optimized with respect to both low rank and sparsity constraints to suppress the noise entries mixed in the exemplar.
\section{Proposed Speech Enhancement Method}
\subsection{Model}
RPCA related approaches generally decompose an input spectrogram matrix \(Y\in \mathbb{R}^{N\times M}\) as the summation of a low-rank matrix \(L\) reflecting less spectrally diverse noise, and a sparse matrix \(S\), representing the sparsity of speech energy as shown in (1). Outlying entries \(E\) are frequently added to provide a relaxed estimation on the noise residuals \cite{sun2014novel, bouzid2016speech}. 
\begin{equation}
\begin{split}
 \min_{S,L} \qquad  &\| S\|_{1}+\gamma_{l}\|L\|_{*}+\gamma_{e}\|E\|_{F}^{2}\\
 s.t.  \qquad &Y  = S+L+E
\end{split}
\end{equation}
$\|\cdot\|_{*} $ is the nuclear norm of the matrix. The sparsity of $S$ is measured by the $L_{1}$ norm $\|\cdot \|_{1}$, and the Frobenious norm is used to represent the noise residual. Based on the observation that speech spectrogram is sparsity in short time period but with repeated structures (i.e., low-rank) within several consecutive segments, \(S\) in (1) can be replaced by \( Y_{1}S\), where as an explicit dictionary of speech spectral templates, \(Y_{1}\) is multiplied by a set of temporal activation \(S\). Generally, \(Y_{1}\) is a global offline trained dictionary \cite{chen2013speech}. However, such kind of final learned dictionary may lose focus on the intermediate estimation of the local speech, and accordingly, is not necessarily good enough to explain current speech frame. Instead, an online speech exemplar can be estimated based on SPP given by
\begin{equation}
Y_{1} = Y \odot \mathcal{P}
\end{equation} 
$\odot$ refers to element-wise product of matrices, and $\mathcal{P}$ is the SPP,  for each frequency bin defined as
\begin{equation}
\begin{split}
\mathcal{P}_{j}  &= p\left\lbrace z_{1} \mid y_{j}, \lambda \right\rbrace \\ 
 &= \frac{p(y_{j} \mid z_{1}, \lambda)p(z_{1})}{\sum_{n=0}^{n=1} p(y_{j} \mid z_{n}, \lambda)p(z_{n})} 
= \frac{p(y_{j} \mid z_{1}, \lambda)w_{1}}{\sum_{n=0}^{n=1} p(y_{j} \mid z_{n}, \lambda)w_{n}}
\end{split}
\end{equation}
where index $n\in \{1,0\}$ represents the case of speech present and absent, respectively. $w_{n}$ is the corresponding the $a$ $priori$ probability that $(\sum_{n}w_{n}=1)$. $\lambda \triangleq \{\mu_{n}, \sigma_{n}, w_{n}\}$ is the parameter set (i.e., mean, variance, and priori) for the statistic models of speech and noise. Considering noise obeys complex Gaussian distribution, $p(y_{j}\mid z_{0}, \lambda)$ should be Rayleigh distribution, and specifically, $p(y_{j}\mid z_{1}, \lambda)$ is assumed as generalized Gamma distribution given by \cite{hendriks2007map}
\begin{equation}
p(y_{j} \mid z_{0},\lambda) = \frac{y_{j}}{\delta^{2}}exp\{-y_{j}^{2}/2\delta^{2}\}
\end{equation}
\begin{equation}
p(y_{j} \mid z_{1},\lambda) = \frac{\beta^{\nu}}{\Gamma(\nu)}y_{j}^{\nu-1}exp\{-\beta y_{j}\}
\end{equation} 
where 
\begin{equation}
\delta = \mu_{0}\sqrt{2/\pi};
\beta = \mu_{1}/\sigma_{1}^{2};
\nu= \mu_{1}^{2}/\sigma_{1}^{2}
\end{equation}

Usually, the local exemplar \(Y_{1}\) promises a sparse activation matrix \(S\) as the representation of individual speech segment from the same subspace. However, in low SNR scenario, the estimated local exemplar will be highly corrupted by noise components, and such kind of speech 'bases' is often overcomplete. Hence there can be many feasible solutions to \(Y=Y_{1}S+L+E\). To address this issue, both the most sparse and lowest rank criteria are imposed on \(S\). The main consideration is that in the estimated dictionary, speech components have larger or at least comparable magnitudes than noise entries, and the low-rank constraint imposed on \(S\) obviously utilizes the similar frequency structures in speech spectral bases. Comparatively, the residual noises in the estimated dictionary can be regarded as sparse components. Thus, we seek a representation \(S\) by solving the following optimization problem
\begin{equation}
\begin{split}
\min_{S,L} \qquad &\|S\|_{*}+\beta \|A\|_{1}+\gamma_{l} \|L\|_{*}+\gamma_{e}\|E\|_{2,1}\\
s.t. \qquad &Y = Y_{1}S+L+E, S=A.
\end{split}
\end{equation}
where an auxiliary variable \(A\) is introduced to make the objective function separable. $\|E\|_{2,1}=\sum_{i=1}^{M}\sqrt{\sum_{j=1}^{N}E_{ij}^{2}}$ is called the $\ell_{2,1}$ norm. The $\ell_{2,1}$ norm encourages the columns of \(E\) to be zero, which assumes that the outlying entries are "sample-specific".

\subsection{Algorithm}
For each row of spectrogram matrix, the probability density function (PDF) is given by 
\begin{equation}
p(y \mid \lambda) = \prod_{j=1}^{M}p(y_{j}\mid \lambda)
\end{equation} 
where \(p(y_{j}|\lambda)=\sum_{z_{n}}p(y_{j}|z_{n},\lambda)p(z_{n})\). The parameter set $\lambda$ is estimated by maximizing the above PDF function. The following are the typical EM re-estimation formulas
\begin{equation}
\widehat{w}_{0,n} = \frac{1}{M}\sum_{j=1}^{M}p(z_{n} \mid y_{j}, \lambda_{0}^{'}) 
\end{equation}
\begin{equation}
\widehat{\mu}_{0,n} = \frac{\sum_{j=1}^{M}y_{j}p(z_{n} \mid y_{j}, \lambda_{0}^{'})}{M\widehat{w}_{0,n}}  
\end{equation}
\begin{equation}
\widehat{\sigma}_{0,n}^{2} = \frac{\sum_{j=1}^{M}(y_{j}-\widehat{\mu}_{0,n})^{2}p(z_{n} \mid y_{j}, \lambda_{0}^{'})}{M\widehat{w}_{0,n}}  
\end{equation}
where \(p(z_{n}|y_{j},\lambda_{0}^{'})\) is obtained through (3)-(6) with the old parameter set \(\lambda_{0}^{'}\). $\widehat{\lambda}_{0}\sim \{\widehat{w}_{0,n}, \widehat{\mu}_{0,n}, \widehat{\sigma}_{0,n}^{2}\}$ denotes the new parameter set re-estimated from $\lambda_{0}^{'}$. In the next iteration, $\lambda_{0}^{'}$ is replaced by $\widehat{\lambda}_{0}$. The initial parameters are obtained by K-means, and the speech model parameter \(\nu\) is fixed if the mean values of speech and noise components are too close. This iteration continues until EM algorithm converges. Based on the obtained local SPP, an online updated schematic is proposed \cite{ying2011voice} as described in Algorithm 1.
\begin{algorithm}[h]
\caption{{\bf Online estimated SPP by EM algorithm} \label{LRR}}
{\bfseries Input:}speech spectrogram matrix Y $\in \mathbb{R}^{N\times M}$ \\
{\bfseries Initialize:}Set maxIter, tolerance $\epsilon$, threshold $\theta_{i}$, where Step1-3 is conducted in \(i\)th row, i\(\leq N\).
\begin{algorithmic}[1]
{\item[]\bfseries  Step 1} Pre-classification
\STATE Initialize $\lambda_{0}^{'}=\{\mu_{0,n}^{'}, \sigma_{0,n}^{'},w_{0,n}^{'}\}$ by K-means.
\IF {$|\mu_{0,1}^{'}-\mu_{0,0}^{'}|\leq \theta_{i}$}
     \STATE Fixed \(\nu\);
\ENDIF
\nonumSTATE {\bfseries  Step 2} EM based SPP
\WHILE{$\|\widehat{\lambda}_{0}-\lambda_{0}^{'}\|\geq \epsilon$ or (k $\leq$ maxIter)}
   \STATE E-step : Calculate the expectation PDF in (8) by (3)-(6)
   \STATE M-Step: Calculate the new parameter set \(\widehat{\lambda}_{0}=\{\widehat{w}_{0,n}, \widehat{\mu}_{0,n}, \widehat{\sigma}_{0,n}^{2}\}\) by (9)-(11), and \(\lambda_{0}^{'}\leftarrow \widehat{\lambda}_{0}\).   
\ENDWHILE \\
{\bfseries Step 3} Online updated SPP\\
\STATE Input: \(\widehat{\lambda}_{0}\rightarrow \lambda_{j} (j\leq M)\) and parameter \(\alpha\) \\
\STATE update \(\lambda\) for each new input \(y_{j}\)(\(j>M\))
     \begin{align*}
      w_{j,n} = \alpha w_{j-1,n}+ (1-&\alpha)p(z_{n}\mid y_{j},\lambda_{j-1})\\
       \mu_{j,n} = \alpha\frac{w_{j-1,n}\mu_{j-1,n}}{w_{j,n}} &+ (1-\alpha)\frac{p(z_{n}\mid y_{j},\lambda_{j-1})y_{j}}{w_{j,n}}\\
       \sigma_{j,n}^{2} = \alpha\frac{w_{j-1,n}\sigma_{j-1,n}^{2}}{w_{j,n}}& \\
         +(1-\alpha)&\frac{p(z_{n}\mid y_{j},\lambda_{j-1})(y_{j}-\mu_{j,n})^{2}}{w_{j,n}}
     \end{align*} 
\end{algorithmic}
{\bfseries Output:}SPP matrix $\mathcal{P}_{ij} = p\left\lbrace z_{1} \mid y_{ij}, \lambda \right\rbrace$
\end{algorithm}

Comparing with conventional Gaussian speech model, the generalized Gamma distribution is more restrictive on noise basis spectra and can accurately pick up the speech components \cite{hendriks2013dft}. In addition, the online update scheme allows the parameter set to be highly descriptive on local distribution. To solve (7), a recently developed method called the linearized alternating direction method with adaptive penalty (LADMAP) has been applied to obtain the optimization result \cite{zhuang2012non}. The augmented Lagrangian function of (7) is
\begin{equation}
\begin{split}
&\mathcal{L}(S,A,L,E,\rho,\Delta_{1},\Delta_{2}) \\
&= \|S\|_{*}+\beta\|A\|_{1}+\gamma_{l}\|L\|_{*}+\gamma_{e}\|E\|_{2,1}\\
&+\frac{\rho}{2}\|Y-Y_{1}S-L-E+\frac{\Delta_{1}}{\rho}\|_{F}^{2}+\frac{\rho}{2}\|S-A+\frac{\Delta_{2}}{\rho}\|_{F}^{2}\\         
\end{split}
\end{equation} 
where \(\Delta_{1}\) and \(\Delta_{2}\) are Lagrangian multipliers. With some algebra, the updating schemes are outlined in Algorithm 2.
\begin{algorithm}
\caption{{\bf Proposed model to solve problem (7)} \label{LADMAP}}
{\bfseries Input:}Speech spectrum matrix \(Y\) $\in \mathbb{R}^{N\times M}$, estimated dictionary \(Y_{1}\), parameters $\gamma_{l}>0$ and $\gamma_{e}>0$, $\rho_{0}>0$, and $\mu>1$. \\
{\bfseries Initialize:}Set maxIter, and tolerance $\epsilon$ False. Initialize $S_{0}$, $A_{0}$, $L_{0}$, $E_{0}$ and $\Delta_{0}$ to zero. 
\begin{algorithmic}[1]
\WHILE{$\|Y-Y_{1}S_{k}-L_{k}-E_{k}\|_{F}/\|Y\|_{F} \geq \epsilon$ or $k\leq$ maxIter)}
   \STATE Update $S_{k+1},A_{k+1},L_{k+1},E_{k+1}$: 
        \begin{align*}
        S_{k+1}=&\Theta_{\frac{1}{\eta\rho_{k}}}([Y_{1}^{T}(Y-Y_{1}S_{k}-L_{k}-E_{k}+\frac{\Delta_{1,k}}{\rho_{k}})-\\
        &(S_{k}-A_{k}+\frac{\Delta_{2,k}}{\rho_{k}})]/\eta+S_{k}) \\
        A_{k+1} =& \mathcal{SR}_{\beta\rho_{k}^{-1}}(S_{k+1}+\frac{\Delta_{2,k}}{\rho_{k}}) \\
        L_{k+1} =& \Theta_{\frac{\gamma_{l}}{\rho_{k}}}(Y-Y_{1}C_{k+1}-E_{k}+\frac{\Delta_{1,k}}{\rho_{k}})\\
        E_{k+1} =& \Omega_{\frac{\gamma_{e}}{\rho_{k}}}(Y-Y_{1}S_{k+1}-L_{k+1}+\frac{\Delta_{1,k}}{\rho_{k}})
        \end{align*}
\STATE Update the Lagrangian multipliers:
\begin{align*}
\Delta_{1,k+1} &= \Delta_{1,k+1}+\rho_{k}(Y-Y_{1}S_{k+1}-L_{k+1}-E_{k+1})\\
\Delta_{2,k+1} &= \Delta_{2,k+1}+\rho_{k}(S_{k+1}-A_{k+1})\\
\rho_{k+1} &= \mu\rho_{k}.
\end{align*}
\ENDWHILE
\end{algorithmic}
{\bfseries Output:}Optimal active coefficient matrix $S_{*}=S_{k}$ \\
\vspace{-2mm}
\end{algorithm}
$\Theta$, $\mathcal{SR}$, and $\Omega$ are the singular value thresholding, shrinkage, and the $\ell_{2,1}$ minimization operator, respectively, and $\eta=\|Y_{1}\|_{2}^{2}$ \cite{zhuang2012non}.    

\section{Experimental evaluation}
The noisy speech signals were synthesized by adding speech samples to different types of noises at various input SNRs(i.e., -10, -5, 0, 5, and 10 dB). Thirty speech samples were selected from NOIZEUS database, and 30 were randomly selected from IEEE wide band speech dataset \cite{loizou2013speech}. Nine different noise samples were used, including six noises (i.e., car, babble, airport, exhibition, restaurant, and train) from AURORA database, two simulated noises (i.e., Gaussian and pink noise), and one jackhammer noise sample from \cite{loizou2013speech}. All signals were resampled to 8 kHz sampling rate, and the spectrograms were calculated with a window length of 32 ms, and a hop of 10 ms. The performance of the proposed LDLSD algorithm was evaluated by comparing with four other algorithms, including one conventional subspace approach (i.e., KLT \cite{hu2003generalized}), and three state-of-the-arts (i.e.,MMSE-SPP \cite{hendriks2013dft}, NMF-RPCA \cite{chen2013speech}, and RPCA \cite{lin2010augmented}). In NMF-RPCA algorithm, the speech dictionary \(Y_{1}\) was learned from the spectrograms of all 60 speech utterances used in this study. Sparse NMF with a generalized KL-divergence \cite{chen2013speech} was used to obtain the dictionary, which consisted of 300 bases. In other words, 5 basis vectors were extracted for each speech utterance. 

An intuitive comparison of the improved speech spectrogram by the proposed LDLSD and two RPCA based algorithms has been shown in Fig.1. For NMF-RPCA, the artificial frequency components can be found around 2.7 second (as circled in Fig.1c). It indicates that the global dictionary bases may lead to an overfitting situation. In addition, the overlap of speech and noise basis convex hull can cause speech-similar-structure noise components(as circled around 1 second in Fig.1c). For RPCA, low rank speech spectrum ingredients are very likely to be wrongly decomposed into noise subspace (as circled at the frequency band 0.5-1 kHz in Fig.1f). Comparatively, the proposed LSLSD demonstrates better decomposition results shown in Fig.1g and 1h. The majority of noise components are correctly decomposed into $L$. Moreover, the speech matrix $Y_{1}S$ in Fig.1g includes most low frequency components, and has least signal distortions than the speech matrix obtained by NMF-RPCA and RPCA algorithms in Fig.1c and 1e.    
\begin{figure}[!hbt]
\vspace{-5mm}
\centerline{\includegraphics[scale=0.56]{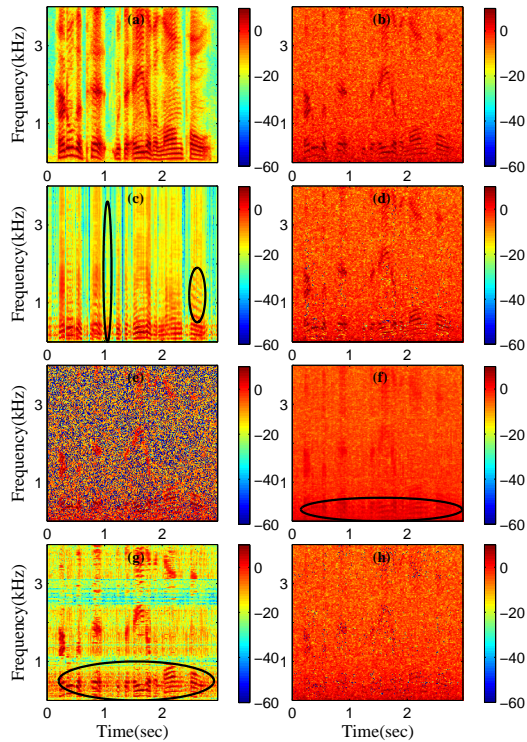}}
\vspace{-4mm}
\caption{The spectrograms of clean speech (a), noisy speech with a pink noise (SNR = 0dB) (b), speech matrixes obtained by NMF-RPCA (c), RPCA (e), and the proposed LDLSD (g) algorithms, and noise matrixes obtained by NMF-RPCA (d), RPCA(f), and LDLSD (h) algorithms.}
\vspace{-6mm}
\end{figure}

Two metrics, signal-to-distortion ratio (SDR) calculated by BSS\_EVAL package and perceptual evaluation of speech quality (PESQ), are used to evaluate speech enhancement algorithms. Figure 2a shows that the averaged SDRs of the enhanced speeches by applying five different algorithms. The proposed LDLSD algorithm demonstrates the highest SDRs at all SNRs (-10, -5, 0, 5, and 10 dB). It indicates that the proposed algorithm can more effectively separate speech components from background noises. In addition, compared with NMF-RPCA, the LDLSD has a comparable averaged SDR at SNR = -10 dB, but demonstrates significantly higher averaged SDRs than the NMF-RPCA at other SNRs (i.e., -5, 0, 5, and 10 dB). 
\begin{figure}[!hbt]\centering
\subfigure{\includegraphics[scale=0.20]{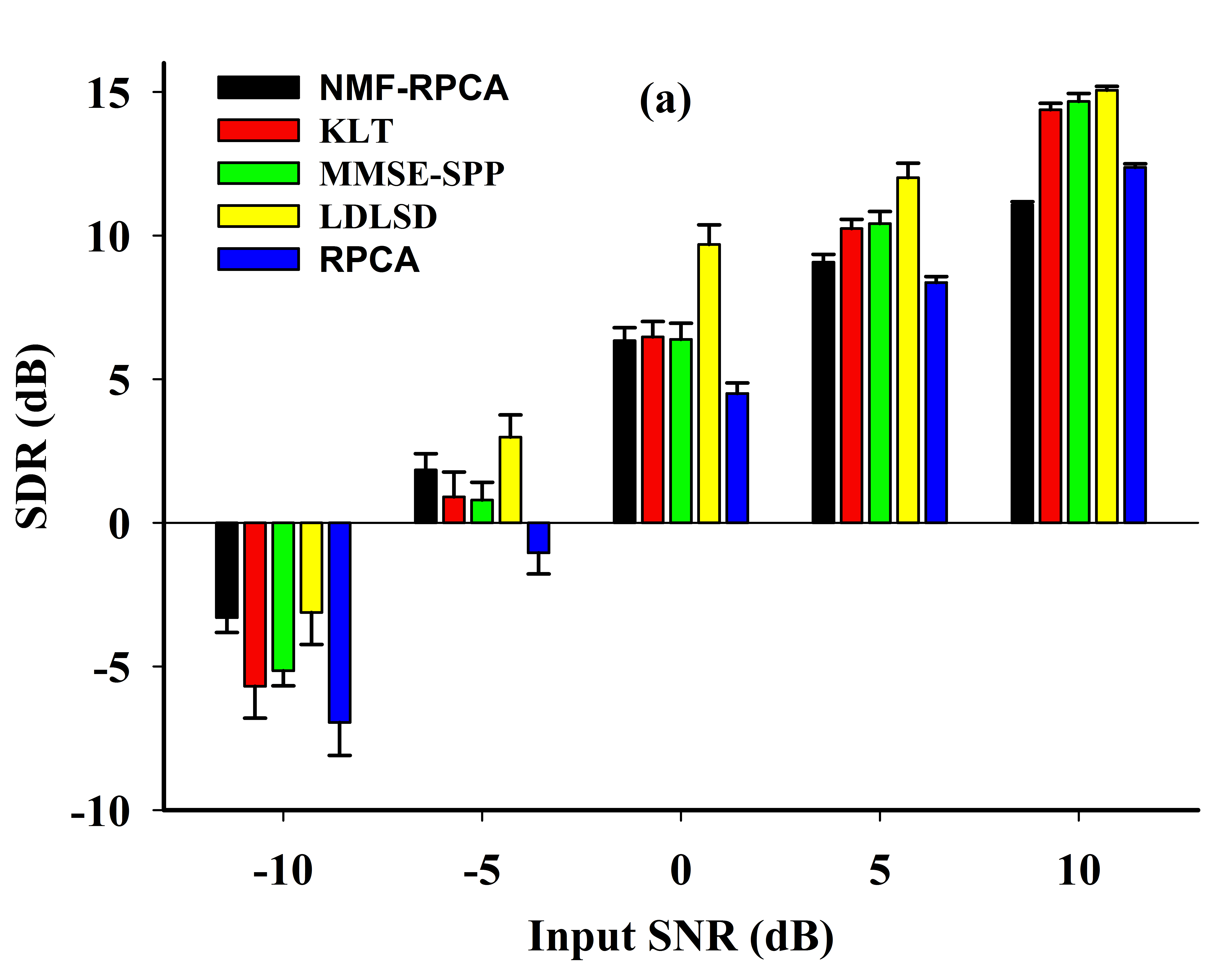}}
\subfigure{\includegraphics[scale=0.20]{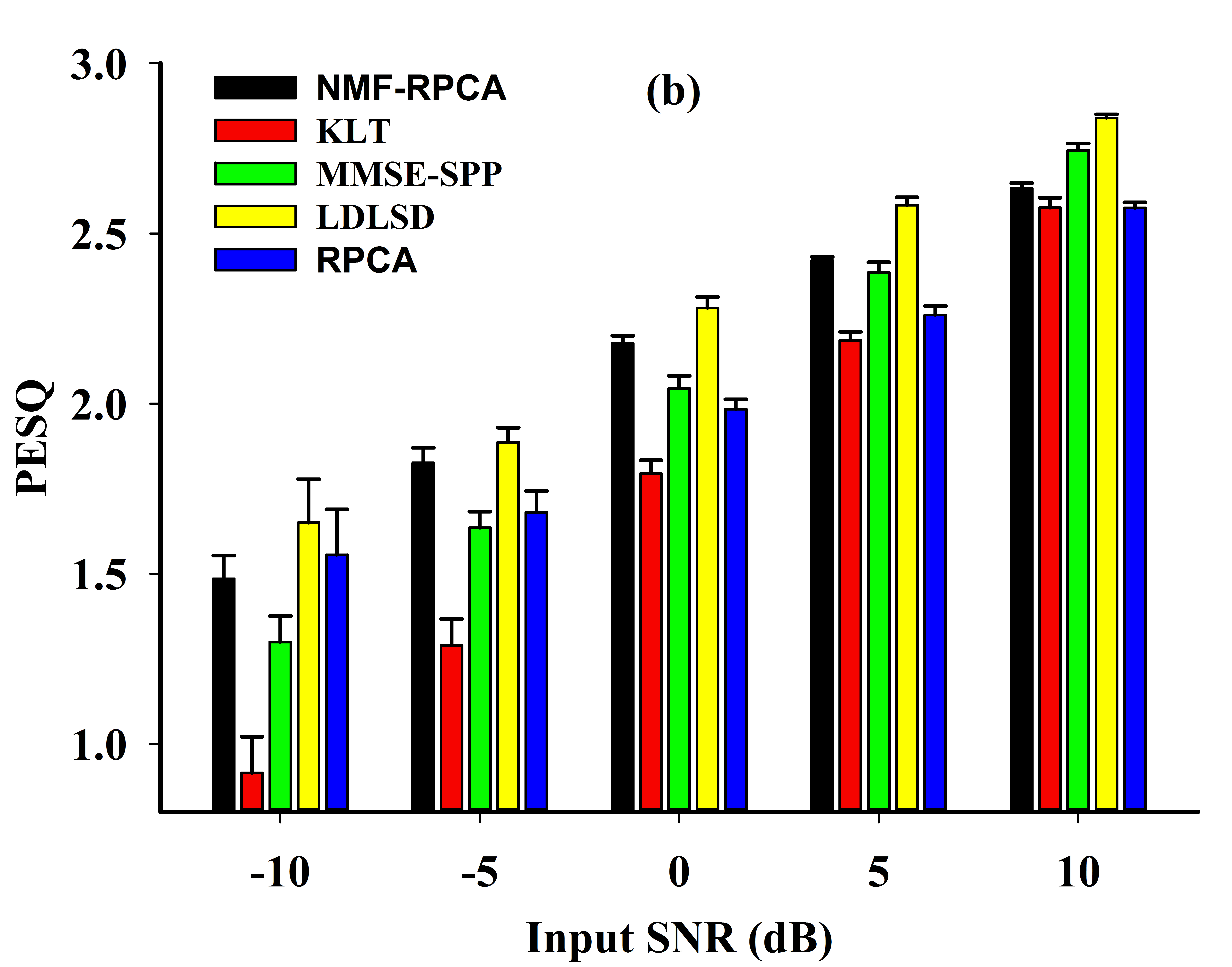}}
\caption{Averaged SDR (a) and PESQ (b) for enhanced speech by applying five algorithms, including NMF-RPCA, KLT, MMSE-SPP, the proposed LDLSD, and RPCA at various SNRs (-10 dB $<$ SNRs $<$ 10 dB), and averaged across eight different types of noise.}
\vspace{-5mm}
\end{figure}

The PESQ scores of enhanced speeches by five algorithms are shown in Fig.2b. The LDLSD algorithm shows significantly higher PESQ improvements than other four algorithms, averagely 0.2 higher than NMF-RPCA, 0.3 higher than RPCA, and 0.4 higher than KLT. Especially at low SNRs (-10 and -5 dB), the low rank and sparse criteria imposed on the activation matrix \(S\) help achieve a better performance than NMF-RPCA algorithm.

In addition, the jackhammer noise, as a highly transient noise, is applied to evaluate the performance of the proposed algorithm. Figure 3 shows the SDRs and PESQ of the enhanced speeches from a jackhammer noise background at various SNRs by five algorithms. The LDLSD demonstrates an obvious advantage over other four algorithms on both two metrics. In magnitude spectra space, the transient noises (e.g., jackhammer noises) and global speech dictionary may be partially overlapped, for example the impulsive components are quite similar to the speech fricatives. This can cause an ambiguity in speech and noise separation in NMF-RPCA. Comparatively, the proposed LDLSD algorithm has two obvious merits: 1) the local estimated exemplar can help to exclude most of the transient features; 2) the low rank constraint imposed on the activation matrix \(S\) can also reduce the impact of transient noise residuals in the online estimated dictionary \(Y_{1}\).
\begin{figure}[!hbt]
\vspace{-4mm}
\centerline{\includegraphics[scale=0.60]{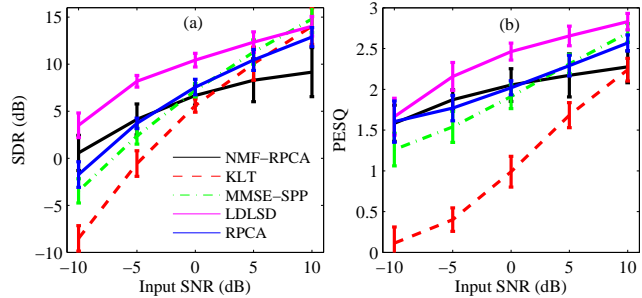}}
\vspace{-2mm}
\caption{Averaged SDR (a) and PESQ (b) for enhanced speech corrupted by transient jackhammer noises, by applying five algorithms.}
\vspace{-4mm}
\end{figure}

\section{Conclusion}
In this letter, we investigate how the SPP based local speech dictionary can be employed in the subspace framework to obtain the low rank and sparse components of speech spectrogram for noisy speech enhancement. A local dictionary based low rank and sparsity decomposition has been proposed to separate the noise and speech components. An online updated EM algorithm is introduced to obtain SPP matrix according to the input noisy speech matrix. By multiplying this SPP matrix element-wise, the broad bases in the speech subspace can be reduced, which consequently improves the accuracy of local speech dictionary. Moreover, the online estimated dictionary is sufficient enough in basis subspace to avoid speech distortion. Specifically, the most sparsity and lowest rank criteria are both imposed to the activation matrix to achieve a noise-resistant decomposition. The results show that LDLSD algorithm obtains significant improvements at various SNRs w.r.t SDR and PESQ, compared with four algorithms, including KLT, MMSE-SPP, NMF-RPCAR and PCA. The future work of this study includes investigation on noise constraints, such as noise variance and noise modeling.  
\ifCLASSOPTIONcaptionsoff
  \newpage
\fi



%


\bibliographystyle{plain}
\bibliography{references}{}

%







\end{document}